\documentstyle[aps,prl,multicol]{revtex}
\input{epsf.sty}
\begin{document}
\title{One-Dimensional Extended States in Partially Disordered Planar Systems}
\author{Shi-Jie Xiong$^{1,2}$, S.N. Evangelou$^{2,3}$, and
E.N. Economou$^{2,4}$ }
\address{$^1$Laboratory of
Solid State Microstructures and Department of Physics, \\
Nanjing University, Nanjing 210093, China \\ 
$^2$Foundation for Research and Technology, 
Institute for Electronic Structure and Lasers,
Heraklion, P.O. Box 1527, 71110 Heraklion, Crete, Greece \\ 
$^3$Department of Physics, University of
Ioannina, Ioannina 451 10, Greece \\  
$^4$Department of Physics, University of
Crete, Heraklion, Greece}  
\maketitle
\begin{abstract}
We obtain analytically  a continuum of one-dimensional ballistic
extended states in a two-dimensional disordered
system, which consists of compactly coupled random and pure
square lattices. The extended states give a marginal metallic 
phase with finite conductivity $\sigma_{0}=2e^2/h$ 
in a wide energy range, whose boundaries define 
the mobility edges of a first-order metal-insulator transition.
We show current-voltage duality, $H_{\parallel }/T$ scaling of
the conductivity in parallel magnetic field $H_{\parallel}$ and
non-Fermi liquid properties when 
long-range electron-electron interactions are included. 

 
\end{abstract}
\vspace{1cm}

\begin{multicols}2
\narrowtext
%\newpage

Anderson localization and the associated metal-insulator transition 
(MIT) remain one of the most active fields in condensed matter 
physics \cite{1}. The one-parameter scaling theory 
predicts localization of all the states  in one or two dimensions 
for any amount of disorder \cite{2}, although even in one-dimension 
($1D$) isolated extended states can exist as a consequence of
specific short-range disorder correlations \cite{3,4}.  
In a two-dimensional ($2D$) disordered system delocalization is 
possible for noninteracting electrons only by breaking the time-reversal 
symmetry via a magnetic field \cite{5} or the spin-rotation 
invariance by spin-orbit coupling \cite{6}.  Extended 
states can also appear  in continuous $2D$ models
with random $\delta $-function potentials 
of infinitesimal action distance \cite{7}. 
The presence of extended states in $2D$ random systems
is of great interest since metallic phase and  
MIT has been recently reported in Si-MOS 
planar structures \cite{8,9,10} 
and GaAs-based materials \cite{11}, in the absence 
of magnetic fields. 

We analytically show extended  wave functions 
in partially random structure 
composed of two compactly coupled square lattices, 
one random and the other periodic.  
At the same time, according to the scaling theory the 
rest of the  states are localized. 
Moreover, the extended states have special momentum and form 
a $1D$ continuum which gives  $2e^{2}/h$ conductivity, 
in a finite energy range. The  boundaries of this range
define the mobility edges (ME) of a first-order MIT. 
These states are perfect (ballistic) and coexist with localized states 
in their energy range. They correspond to a non-Fermi liquid, 
in agreement with recent scaling expectations for the 
presence of a MIT in $2D$ \cite{12}. We show that
the system  under  long-range Coulomb interactions 
reduces to the marginal Fermi liquid proposed in \cite{13}. 

We consider two coupled  
square lattices with the sites of one  located above the plaquette centers 
of the other. One plane is pure and the other is random.  
The tight-binding Hamiltonian  reads
\begin{eqnarray}
\label{ham}
H & = & \sum_{\lambda ,i} \epsilon_{\lambda ,i} 
a^{\dagger}_{\lambda ,i}a_{\lambda ,i} +
\sum_{\lambda ,\langle i,j \rangle }t_{\lambda} (a^{\dagger }
_{\lambda ,i}a_{\lambda ,j} +\text{H.c.} ) \nonumber \\
& + & \sum_{\langle 1i, 2j \rangle }t(a^{\dagger }
_{1,i}a_{2,j} +\text{H.c.} ) ,
\end{eqnarray}
where $a_{\lambda ,i}$ is the destruction operator for an electron 
at site $i$ of the $\lambda =1(2) $ pure (random) plane, 
$t_{\lambda}$ is the nearest-neighbor (NN) hoppings within the $\lambda$-th 
plane and the third term connects each random site to its 
NN pure sites via the hopping parameter $t$. We choose 
$t_1=t_2=1$, setting the energy unit throughout, 
although our conclusions are also valid if $t_2$ is random.  
The sites of the pure plane 
have energies $\epsilon_{1,i}=0$ and of the random 
plane $\epsilon_{2,i}$, randomly distributed between 
$-W/2$ and $W/2$. For  convenience, the length units 
and the origin of the coordinates are set so that 
the pure (random) sites count odd (even) numbers. 
 
Obviously, extended wave functions can have 
zero amplitude on the random sites
and finite amplitude on the pure sites, being
not influenced by randomness. 
We can seek for such states in the considered 
partially disordered system. 
Indeed, for an $L \times L$ lattice with 
the described geometry and periodic boundary conditions
in all directions,
with $L$ even and $x_n$, $y_n$ the coordinates of the 
$n\equiv (\lambda ,i)$-th site,  the states
\begin{equation}
\label{ext1}
\psi_{1,k_x}(x_n,y_n)=\frac{2}{L}\sin (\frac{y_n\pi }{2})\exp (ik_x x_n),
\end{equation}
\begin{equation}
\label{ext2}
\psi_{2,k_y}(x_n,y_n)=\frac{2}{L}\sin (\frac{x_n\pi }{2})\exp (ik_y y_n), 
\end{equation} 
where $k_{x}(k_{y})=2j_{x}(j_{y})\pi/L, 
j_{x}(j_{y})=1,2,...,L/2$, are exact eigenstates of 
the Hamiltonian $H$ having zero amplitude on the random sites.
This can be easily verified by applying $H$ on 
$\psi_{1,k_x}(x_n,y_n)$, $\psi_{2,k_y}(x_n,y_n)$
to obtain their eigenenergies
\begin{equation}
E=2[\cos (2k_x)-1] \text{ and } E=2[\cos (2k_y)-1],
\end{equation}
respectively.  The transverse momentum of these states 
is fixed to  $k_{y}(k_{x})=\pi/2$ and their longitudinal momentum 
$k_{x}(k_{y})$ runs in a $1D$ Brillouin zone, forming a $1D$ continuum 
in the energy range $[-4,0]$ \cite{14}.

\begin{figure}[tbh]
\epsfxsize=3in
\epsfysize=3in
\epsffile{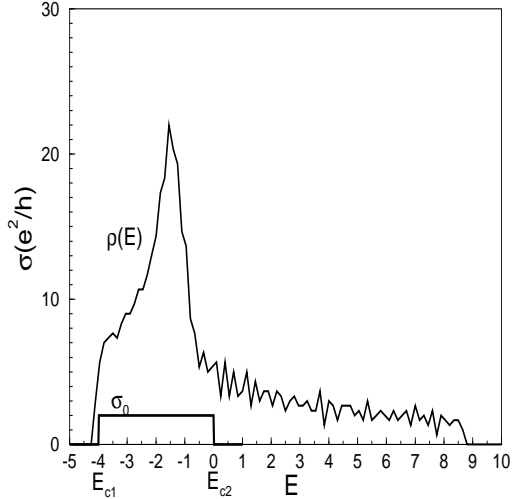}
\vspace{3mm}
\caption[fig1]{\label{fig1} The density of states $\rho(E)$,
in arbitrary units, for a finite $2D$ random sublattice system of 
linear size $L=52$ with $t=1.2$ and $W=1.5$ by taking
averages over 100 random configurations. The scale invariant
conductivity $\sigma_0$ is also shown.}
\end{figure}

The perfectly extended states of Eqs. (\ref{ext1}) and (\ref{ext2}) 
have momentum along both the $2D$ principal axes and
do not violate the scaling theory, 
since they effectively decouple from the random sites
although the two sublattices are compactly coupled.
At a given energy $E$, with $-4 \leq E \leq 0$, they 
provide propagating channels with $\sigma_0=2 e^2/h$ 
conductivity (if spin is included) along the $x$ (or $y$) direction,  
independently of the system size $L$. 
In Fig. 1 we plot $\sigma_0$  and the density of states $\rho(E)$ 
as a function of  $E$.  The rest of the $2D$ states are
asymptotically localized, as expected from the scaling theory \cite{2}. 
The transparent channels
give minimum metallic conductivity $\sigma_{0}=2 e^2/h$
and  one obtains a first-order MIT from $\sigma_{0}$ to $0$ at 
$E_{c1}=-4$ and $E_{c2}=0$, which can be regarded as 
mobility edges. For the chosen structure we observe 
that the lower ME lies near the 
bottom of the band (see Fig. (1)), so that the  
Fermi energy can move into the 
metallic regime even for small electronic doping. 
For example, this is the case 
in silicon MOSFETs where the metallic phase is reached at 
very low electron densities ($\sim 10^{11} $cm$^{-2}$) \cite{8,9,10}. 
It should be also noted that a minimum metallic conductivity
close to $e^2/h$ has been recently suggested,
from experimental data in $2D$ hole systems 
of GaAs heterostructures \cite{15}.   

The obtained conductivity $\sigma=\sigma_0$ is 
invariant under scaling so that
from the viewpoint of the scaling theory it 
does not imply  a {\it true} metallic phase for the system,
but a critical phase instead, with the corresponding 
$\beta$-function only reaching
zero \cite{2}. Moreover, since the extended states coexist 
with localized states in 
a wide energy range, having lower dimensionality and measure, 
we can rename the obtained phase ``marginal metallic'', 
following the terminology ``marginal Fermi liquid'' 
proposed by Varma {\it el al} \cite{13}. 

\begin{figure}[tbh]
\epsfxsize=3in
\epsfysize=3in
\epsffile{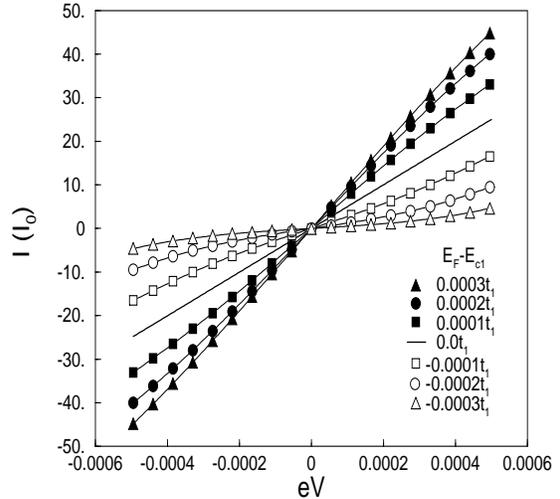}
\vspace{3mm}
\caption[fig2]{\label{fig2} The current-voltage curves at several 
Fermi levels around the mobility edge with temperature $k_{B}T=
1.1 \times 10^{-5} t_{1}$.}
\end{figure}

The step-like conductivity $\sigma (E)=\frac{2e^2}{h}\theta (E-E_{c1}) $ 
at $E_{c1}$ gives nonlinearity of the current--
voltage ($I-V$) curves with $I-V$ duality.  
If the sample is connected to two metallic leads 
with the same Fermi level $E_{F}$, 
an applied voltage $V$ between the two contacts opens 
a transmission window of width $eV$ around $E_{F}$ at
zero temperature, due to the 
electronic occupation in the leads.  
Nonlinearity occurs if the window boundary passes 
across the ME when changing the voltage. 
It becomes smoother at finite temperature $T$.  
In order to illustrate this effect we compute the current 
from  the formula
\begin{eqnarray}
I(V) & = & {\frac {I_0}{k_{B}T}}\int_{E_{c1}}^{E_{c_2}}\left( \frac{1}
{1+\exp [(E-E_F-eV/2)/k_BT]} \right. \nonumber \\
& - & \left. \frac{1}{1+\exp [(E-E_F+eV/2)/k_BT]}\right) dE ,
\end{eqnarray}
where $I_0$ is a prefactor.
In Fig. 2 the calculated $I-V$ curves 
display nonlinearity if $E_{F} \neq E_{c1}$. 
For $E_F$ larger (smaller) than $E_{c1}$ the resistivity increases (decreases) 
with $|V|$, giving the symmetric behavior about 
$E_F=E_{c1}$ known  as $I-V$ duality, exchanging the insulating 
with  the metallic phases. Similar kind of duality has been  
observed in silicon MOSFETs (Fig. 1 of Ref. \cite{10}).

The extended electrons, decoupled from the rest, 
form a pseudo-Fermi sea embedded in 
the localized electrons. The pseudo-Fermi surface has only four 
points corresponding to two 
Luttinger liquids crossing with each other.
This  provides a possibility 
for applying the $1D$ or quasi-$1D$ Luttinger liquid theory to $2D$ 
systems, as pointed out several years ago \cite{16}. 
Nevertheless, the system should be regarded as $2D$ non-Fermi liquid owing 
to its measure, its $1D$-like spectrum, as well as 
the presence of the background from the localized electrons. 

If Coulomb interactions are included
the corresponding matrix elements read
\begin{eqnarray}
\label{inter}
\langle \mu_1,\mu_2 |V|\mu_3,\mu_4\rangle & = & \sum_{{\bf n}_{1},{\bf n}_{2}} c^*_1({\bf n}_1)c^*_2({\bf n}_2)
V(|{\bf n}_1-{\bf n}_2|) \nonumber \\
c_3({\bf n}_1)c_4({\bf n}_2), 
\end{eqnarray}
where $\mu_i$ is a single-particle state of 
$H$ and $c_i({\bf n}_j)$ its coefficient at site ${\bf n}_j$.  
We are interested in elements 
where one of the initial states ($\mu_3$) is extended and the 
other ($\mu_4$) is localized which describe scattering between extended 
and localized electrons.
The extended states have  amplitudes with  
rapidly changing sign in one direction, 
while the localized states show no phase 
coherence having randomly varying amplitudes.  
Thus, for  long-range Coulomb interactions 
the elements with localized  final states $\mu_1$ and $\mu_2$  
are almost zero, due to cancellations arising  from the 
positive and negative signs of the extended waves. 
This leads to an extremely small probability 
for an extended state to  be scattered 
into a localized state, although the number of localized states is much 
higher \cite{14}. 
If a  final state (e.g., $\mu_1$)
is also extended with the same 
wave symmetry as the initial state
the matrix element is of order $\sim 1$, since the 
product $c^*_1({\bf n}_1)c_3({\bf n}_1)$ eliminates the sign of the extended wave 
coefficients. 
The above discussion implies that the extended states found can survive 
in the presence of long-range Coulomb interactions. 
However, they experience fluctuations from 
particle-hole excitations of the localized 
electrons, which can be described by the polarizability
\begin{equation}
\label{green}
\Pi (i\nu_m )\sim 2k_BT\sum_{i,j,\nu_n}G_0(\epsilon_i,i\nu_n)
G_0(\epsilon_j,i\nu_n+i\nu_m),
\end{equation}
where $G_0$ is the single-particle Green function of the localized states, 
the sums for $i$ and $j$ are over the relevant localized 
states for the given interaction, $\epsilon_i $ is the energy of 
state $i$, and $\nu_n=(2n+1)\pi k_BT $ with $n$ an integer. 
After summing over the Matsubara frequencies and 
assuming constant density of the localized states, one has
\begin{equation}
\text{Im} \Pi (\omega ) \sim \left\{ \begin{array}{l} A_0(\omega /k_BT),
\text{ for } |\omega |< k_BT , \\ 
A_0 \text{sign} \omega, \text{ for } |\omega |> k_BT \end{array} 
\right. ,
\end{equation}
where $A_0$ is a prefactor dependent of density of states. 
This is the basic hypothesis of the marginal Fermi liquid theory \cite{13}. 
The key point in this derivation is the 
momentum independence of the Green function for the localized states.     
If there are no added impurities in the 
host sublattice, this manifold is a pure marginal Fermi liquid 
which can give finite conductivity at zero temperature, 
distinct from the non-pure marginal Fermi liquid which experiences 
further impurity scatterings \cite{17}.

\begin{figure}[tbh]
\epsfxsize=3in
\epsfysize=3in
\epsffile{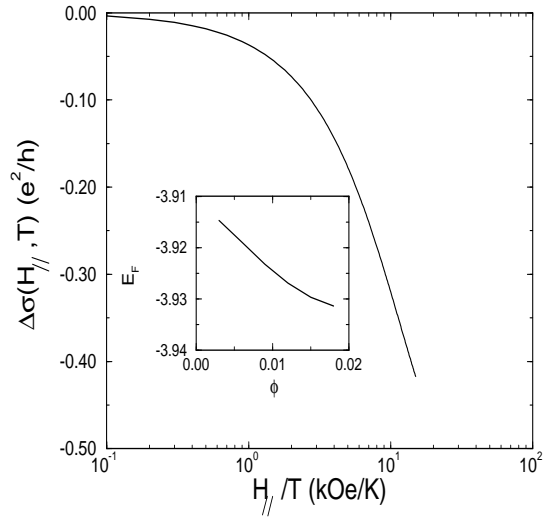}
\vspace{3mm}
\caption[fig3]{\label{fig3}
Main figure: the $H_{\parallel }/T$ scaling  of the conductivity change due to 
a parallel magnetic field for the system at the transition point.  
Inset: the Fermi energy 
$E_F$ vs. $\phi $, the flux through a primary inter-plane triangle 
with the field along the $x$ axis, averaged over $100$ random 
samples with size $L=30$ ($2\times 15 \times 15 $), 
and the rest of parameters as in Fig. 1. 
The units shown are derived for a sample with $a=b=6$\AA, $t_1=1$eV.  
The slope of the inset curve gives $g_1=-1.1 \mu_B$. } 
\end{figure} 

An  in-plane magnetic field applied to the quasi-$2D$ system  
in the Coulomb gauge gives in-plane NN hoppings 
$t_1^{(ij)}=\exp (i\phi \sin \alpha_{ij}) $ and 
$t_2^{(ij)}=\exp (-i\phi \sin \alpha_{ij})$, 
where $\alpha_{ij}$ is the angle between 
the hopping bond $ij$ and the field, 
$\phi =\case{1}{2}H_{\parallel }ab$ is the
flux through a primary inter-plane triangle, $a$ the planar lattice spacing 
and $b$ the distance between the planes. The  inter-plane
NN hoppings are unchanged in this gauge. Thus, the states of 
Eqs. (\ref{ext1}) and (\ref{ext2}) survive  with 
eigenenergies $E=2[\cos (2k_x+2\phi \sin \alpha_x)-\cos (2\phi \sin 
\alpha_y)]$ and $E=2[\cos (2k_y+2\phi \sin \alpha_y)-\cos 
(2\phi \sin \alpha_x)]$, respectively,  where $\alpha_{x(y)}$ is the 
angle between the $x(y)$ axis and the field. 
It can be seen that the shift of $E_{c1}$ (or $E_{c2}$) is of the order of 
$H_{\parallel}^2$ for small fields.  
At the same time, the whole band including 
the localized states changes, so that the Fermi level $E_F$ 
is also shifted if the electron density is kept fixed. 
In order to obtain the field dependence of $E_F$  we compute the energy 
spectrum of a finite random system and the result is 
shown in the inset of Fig. 3. 
In general,  $\Delta (H_{\parallel})$, 
the energy difference between $E_{c1}(H_{\parallel})$ and 
$E_F(H_{\parallel})$, can be expanded as a power series of the field, via
$\Delta (H_{\parallel}) =\Delta (0) +g_1 H_{\parallel}+...$. For small 
fields we can only keep the linear term. 
On the other hand, the spin degeneracy is removed by
$\pm \mu_BH_{\parallel} $ for 
up and down spins, with $\mu_B$ being the Bohr magneton. 
For a small bias voltage the conductivity can be expressed in terms of 
Landauer-B\"{u}tticker formula \cite{18} and 
at the MIT ($\Delta (0)=0$) one has 
\begin{equation}
\sigma (H_{\parallel},T)- \sigma (0,T)
=\frac{e^2}{h}\sum_{g=\pm 1}
\frac{1}{1+\exp \frac{(g_1+g \mu_B)H_{\parallel }}
{k_BT}}, 
\end{equation}
which shows $H_{\parallel}/T$ scaling behavior.  
In Fig. 3 the $H_{\parallel}/T$ dependence of conductivity  
is plotted for one sample, showing good agreement 
with Fig. 3 of Ref. \cite{19}). 

In summary, we have rigorously shown a band of
$1D$ ballistic extended states which propagate freely 
in a $2D$ mixed pure-random system, without  
magnetic field or spin-orbit coupling. 
The $1D$ extended continuum is located at the edges of
the first Brillouin zone ($k_x=\pi/2$ and any $k_y$ or
$k_y=\pi/2$ and any $k_x$) of the $2D$ square lattice.
Moreover, the $1D$ extended  states coexist with localized states, 
in a range of energies whose boundaries define
a first-order MIT. We derive $I-V$ 
duality, $H_{\parallel}/T$ scaling, and demonstrate
non-Fermi or marginal Fermi liquid behavior.  These features 
have not been obtained previously in the presence of disorder
and are of profound significance in understanding 
many experimental findings, especially the $2D$ MIT and 
the underdoped cuprates. 
In this Letter we only focus on the general characteristics 
of  a class of partially disordered materials, with at least one 
sublattice perfect.
In Si MOSFETs and GaAs-AlGaAs heterostructures, 
electrons or holes are confined within 
several atomic planes and due to  lattice mismatch, or atomic diffusion, 
the planes adjacent to the interface are random, while some 
planes relatively far  from the interface are 
less random and may support ballistic channels. 
Another example presents YBa$_2$Cu$_3$O$_{6+x}$ 
where the CuO$_2$ planes are perfect and the CuO$_x$ planes 
random due to oxygen vacancies. 
Although for these materials more realistic models are clearly 
needed, possibly by including more layers, our results may still 
shed light on many of their common features.  

{\bf Acknowledgments.} 
This work was supported in part  by TMR and also a 
$\Pi$ENE$\Delta$ Research Grant of the Greek Secretariat of Science and 
Technology. 

\end{multicols}

\begin{references}
\bibitem{14} We have also performed 
exact diagonalization of the $L\times L$ 
($L/2 \times L/2$ pure and 
$L/2 \times L/2$ random) system with periodic boundary
conditions, which showed  
$L/2+L/2-1 = L-1$ perfect $1D$ extended 
states lying among localized states in the energy regime $[-4,0]$, 
and transfer matrix products which gave exactly zero  
smallest Lyapunov exponent (inverse localization length) 
in the extended regime, for any $L$, $W$.
\end{references}
\end{document}